\documentclass[graybox]{svmult}

\usepackage{mathptmx}
\usepackage{helvet}
\usepackage{courier}

\usepackage{graphicx}
\usepackage{multicol}
\usepackage[bottom]{footmisc}

\usepackage{amssymb}
\usepackage{amsmath}

\usepackage[sort&compress,numbers]{natbib}

\newcommand{\bm}{\mathbf}

\def\fds{0.4}

\newcommand{\dd}{\operatorname{d}\!}
\newcommand{\ee}{\text{e}}

\renewcommand{\eqref}[1]{Eq.~(\ref{#1})}
\newcommand{\intxt}{\int\!\dd\tau\dd\bm x\ }
\newcommand{\intxtp}{\int\!\dd\tau\dd\bm x\dd\tau'\dd\bm x'\ }
\newcommand{\tr}{\operatorname{Tr}}

\newcommand{\fld}[2]{#1_{#2}^{\phantom{*}}}
\newcommand{\flds}[2]{#1_{#2}^{*}}
\newcommand{\bmx}{\bm x}

\begin{document}

\title*{\textsf{Thermodynamics of Trapped Imbalanced Fermi Gases at Unitarity}}

\author{J.M. Diederix \and H.T.C. Stoof}

\institute{J.M. Diederix and H.T.C. Stoof\at Institute for Theoretical Physics, Utrecht University,\\
Leuvenlaan 4, 3584 CE Utrecht, The Netherlands,
\\\email{J.M.Diederix@uu.nl}
}

\maketitle

\abstract{
We present a theory for the low-temperature properties of a
resonantly interacting Fermi mixture in a trap, that goes
beyond the local-density approximation. The theory corresponds
essentially to a Landau-Ginzburg-like approach that includes
self-energy effects to account for the strong interactions at
unitarity. We show diagrammatically how these self-energy
effects arise from fluctuations in the superfluid order
parameter. Gradient terms of the order parameter are included
to account for inhomogeneities. This approach incorporates the
state-of-the-art knowledge of the homogeneous mixture with a
population imbalance exactly and gives good agreement with the
experimental density profiles of Shin {\it et al}.\ [Nature
{\bf 451}, 689 (2008)]. This allows us to calculate the
universal surface tension of the interface between the
equal-density superfluid and the partially polarized normal
state of the mixture. We also discuss the possibility of a
metastable state to explain the deformation of the superfluid
core that is seen in the experiment of Partridge {\it et al}.\
[Science {\bf 311}, 503 (2006)].}

\section{Introduction}

Ultracold atom experiments are always performed in a trap to
avoid contact of the atoms with the `hot' material walls that
would heat up the cloud. Due to this trapping potential the
atomic cloud is never homogeneous. However, typically the
energy splitting of the trap corresponds to a small energy
scale, so that the inhomogeneity is not very severe. In this
case, we may use the so-called local-density approximation
(LDA). It physically implies that the gas is considered to be
locally homogeneous everywhere in the trap. The density profile
of the gas is then fully determined by the condition of
chemical equilibrium, which causes the edge of the cloud to
follow an equipotential surface of the trap.

But even if the trap frequency is small, the LDA may still
break down. An important example occurs when an interface is
present in the trap due to a first-order phase transition. For
a resonantly interacting Fermi mixture with a population
imbalance in its two spin states
\cite{zwierlein2006fsi,hulet2006pps}, such interfaces were
encountered in the experiments by Partridge {\it et al}.\
\cite{hulet2006pps} and by Shin {\it et al}.\
\cite{shin2008pdt} at sufficiently low temperatures. Here the
LDA predicts the occurrence of a discontinuity in the density
profiles of the two spin states, which cost an infinite amount
of energy when gradient terms are taken into account.
Experimental profiles are therefore never truly discontinuous,
but are always smeared out. An important goal of this chapter
is to address this interesting effect, which amounts to solving
a strongly interacting many-body problem beyond the LDA. Due to
the rich physics of the interface, new phases can be stabilized
that are thermodynamically unstable in the bulk. This exciting
aspect shares similarities with the physics of superfluid
helium-3 in a confined geometry \cite{fetter1976thm} and spin
textures at the edge of a quantum Hall ferromagnet
\cite{karlhede1996teq}.

Note that the presence of an interface also can have further
consequences. Namely, in a very elongated trap, Partridge {\it
et al}. observed a strong deformation of the minority cloud at
their lowest temperatures. At higher temperatures the shape of
the atomic clouds still followed the equipotential surfaces of
the trap \cite{partridge2006dtf}. A possible interpretation of
these results is that only for temperatures sufficiently far
below the tricritical point
\cite{shin2008pdt,sarma1963oiu,combescot2004ltf,parish2007ftp,gubbels2006spt,gubbels2008rgt},
the gas shows a phase separation between a balanced superfluid
in the center of the trap and a fully polarized normal shell
around this core. The superfluid core is consequently deformed
from the trap shape due to the surface tension of the interface
between the two phases
\cite{partridge2006dtf,silva2006stu,haque2007tfc}. This causes
an even more dramatic breakdown of the LDA. Although the above
interpretation leads to a good agreement with the experiments
of Partridge {\it et al}.\ \cite{partridge2006dtf}, a
microscopic understanding of the value of the surface tension
required to explain the observed deformations has still not
been obtained. Presumably closely related to this issue are a
number of fundamental differences with the study by Shin {\it
et al}.\ \cite{shin2008pdt}. Most importantly, the latter
observes no deformation and finds a substantially lower
critical polarization, which agrees with Monte Carlo
calculations combined with a LDA. It appears that the
interfaces between the superfluid core and the normal state are
fundamentally different for the two experiments, which might
play an important role in resolving the remaining
discrepancies. In order to investigate this interface we need
to go beyond the local-density approximation.

To study the details of the superfluid-normal interface we need
a theory that can describe the inhomogeneous and population
imbalanced unitarity Fermi gas. For this, we first need a
theory that includes in the homogeneous case both the normal
state and the superfluid state in one quantitative correct
description. Secondly, we need to incorporate the inhomogeneous
effects of the trapping potential. The aim of this chapter is
to give a simple and elegant way to achieve this. In the
following two sections, we fist arrive at an accurate, and to a
large extent analytical description of the thermodynamics of a
population imbalanced unitarity Fermi gas. This is achieved by
constructing an appropriate thermodynamic potential $\Omega$
for the Fermi mixture at unitarity. All desired thermodynamic
quantities can then be obtained by performing the appropriate
differentiations of the thermodynamic potential that are well
known from statistical physics. The inhomogeneity effects of
the trapping potential are included by taking the energy
penalty for large variations in the order parameter into
account. These gradient terms smoothen the jump of the order
parameter that is predicted by the LDA at the location of the
first-order phase transition. We will see that this gives a
more detailed explanation of the experimental data of Shin {\it
et al}.\ \cite{shin2008pdt}. In the last section we then show
how the surface tension can be computed with this more detailed
description of the interface. This surface tension turns out to
be relatively small. This does, therefore, not explain the
dramatic deformation seen in the experiment of Partridge {\it
et al}.\ \cite{partridge2006dtf}. An alternative explanation
may be that there exists a metastable state with a deformed
superfluid core \cite{baksmaty2010cms}. At the end of this
chapter we briefly discuss this possibility. We find that the
Landau-Ginzburg-like theory derived here does not appear to
contain such a metastable state.

\section{Ultracold Quantum Fields}\label{sec:TTI:LG}

In order to properly study the unitary Fermi mixture, we derive
a single thermodynamic potential that in a quantitative correct
manner describes both the normal and the superfluid phases. As
we will see, the normal state of the unitarity Fermi mixture is
straightforwardly incorporated by introducing two
mean-field-like self-energies. In particular, it is possible in
this manner to completely reproduce the equation of state known
from Monte-Carlo calculations. However, including also the
possibility of superfluidity at low temperatures and low
polarizations turns out to be more difficult. To understand
better how this can nevertheless be achieved, we first give an
exact diagrammatic discussion of the superfluid state that is
then in the next section used to arrive at the desired
thermodynamic potential of the unitarity Fermi mixture.

\subsection{Bardeen-Cooper-Schrieffer Theory}

In this subsection we outline the basic ingredients of a
field-theoretical description for the superfluid state of the
imbalanced Fermi mixture \cite{stoof2009uqf}. We start with the
essentially exact action for such an atomic two-component
mixture,
\begin{align}\begin{split}
    S[\phi^*,\phi;J^*, J]  = & \sum_{\sigma=\pm} \intxt\flds{\phi}{\sigma}(\bm x,\tau)\left(\hbar\frac{\partial}{\partial\tau}-\frac{\hbar^2\nabla^2}{2m}-\mu_{\sigma}\right)\fld{\phi}{\sigma}(\bm x,\tau)\\
       & + \intxt V_0\flds{\phi}{+}(\bm x,\tau)\flds{\phi}{-}(\bm x,\tau)\fld{\phi}{-}(\bm x,\tau)\fld{\phi}{+}(\bm x,\tau) \\
       & - \hbar \sum_{\sigma=\pm}\intxt \left(\flds{J}{\sigma}(\bm x,\tau)\fld{\phi}{\sigma}(\bm x,\tau) +\flds{\phi}{\sigma}(\bm x,\tau)\fld{J}{\sigma}(\bm x,\tau) \right)\;.
\end{split}
\end{align}
Here $\fld{\phi}{\sigma}$ is the fermion field of the atomic
species in the hyperfine state $|\sigma\rangle$, $\mu_\sigma$
is the associated chemical potential, $\fld{J}{\sigma}$ is a
Grassmann-valued current source that is convenient in the
following, but which is put equal to zero at the end of the
calculations, and $V_0$ is the strength of the
unitarity-limited attractive interactions between the two
species. The grand-canonical partition function is then given
by
\begin{equation}
    Z[J,J^*] = \int\!\prod_{\sigma}\dd[\flds{\phi}{\sigma}]\dd[\fld{\phi}{\sigma}]\ \exp\left\{-\frac{1}{\hbar}S[\phi^*,\phi;J^*\;,
    J]\right\}.
\end{equation}
This represents a functional integral over all the fermion
fields that are antiperiodic on the imaginary time interval
$[0,\hbar\beta]$, with $\beta = 1/k_{\text{B}}T$ the inverse
thermal energy. The thermodynamic potential is ultimately given
in terms of the partition function as
\begin{equation}
    \Omega(\mu_+,\mu_-,T,V) = -\frac{1}{\beta}\log{Z[0,0]}\;,
\end{equation}
with $V$ the total volume of the system. To make the connection
with thermodynamics explicit, we note that the thermodynamic
potential is related to the pressure $p$ of the gas by means of
$\Omega = -pV$.

In order to describe pairing of the fermions, we perform a
Hubbard-Stratonovich transformation to the complex pairing
field $\Delta$. For this field we have that
\begin{equation}\label{eq:TTI:GapEquation}
    \langle\Delta(\bm x, \tau)\rangle = V_0\langle\phi_-(\bm x, \tau)\phi_+(\bm x, \tau)\rangle\;.
\end{equation}
This transformation makes the action quadratic in the fermion
fields. More precisely, we have that
\begin{align}
    S[\Delta^*,\Delta,\phi^*,\phi; J^*, J] = & -\intxt\frac{|\Delta(\bm
    x,\tau)|^2}{V_0} \\
    &   -\hbar\intxtp  \Phi^{\dagger}(\bm x,\tau)\cdot\bm G_{\rm BCS}^{-1}(\bm x,\tau;\bm x',\tau';\Delta)\cdot\Phi(\bm x',\tau') \nonumber \\
    & + \hbar \intxt\left( J^{\dagger}(\bm x,\tau)\cdot\Phi(\bm x,\tau) + \Phi^{\dagger}(\bm x,\tau)\cdot J(\bm x,\tau) \right)\nonumber \;,
\end{align}
where we defined $\Phi^{\dagger} = [\flds{\phi}{+},
\fld{\phi}{-}]$ and $J^{\dagger} = [\flds{J}{+},\fld{J}{-}]$,
which are vectors in a two-dimensional space, known as Nambu
space. In this space the $2\times 2$ Green's function matrix is
given by $\bm G_{\rm BCS}^{-1}(\bm x,\tau;\bm x',\tau';\Delta)
= \bm G_0^{-1}(\bm x,\tau;\bm x',\tau') - \Sigma_{\rm BCS}(\bm
x,\tau;\bm x',\tau') $. The first term in the right-hand side
represents the noninteracting part and is given by
\begin{align}
    \bm G_0^{-1}(\bm x, \tau;\bm x',\tau') = &
                   \begin{bmatrix}
                     G_{0;+}^{-1}(\bm x, \tau;\bm x',\tau')   & 0 \\
                     0 & -G_{0;-}^{-1}(\bm x', \tau';\bm x,\tau) \\
                   \end{bmatrix}\;,
\intertext{%
with $G_{0;\sigma}$ the noninteracting Green's function of
species $\sigma$. The second term corresponds to the BCS
self-energy, which has only off-diagonal terms and reads }
    \hbar\Sigma_{\rm BCS}(\bm x, \tau;\bm x',\tau') = &
                   \begin{bmatrix}
                     0 & \Delta(\bm x, \tau) \\
                     \Delta^*(\bmx,\tau) & 0 \\
                   \end{bmatrix}\delta(\bmx-\bmx')\delta(\tau-\tau')\;.
    \label{eq:TTI:BCSselfEnergy}
\end{align}

The action now only contains quadratic terms in the fermion
fields, which is something we can handle exactly. However, the
tradeoff is an extra functional integral over the $\Delta$
field. Starting with the easy part, we perform the functional
integral over the fermion fields. Since this is a standard
Gaussian integral, we immediately obtain
\begin{align}
\begin{split}
    S^{\text{eff}}[\Delta^*,\Delta;J^*,J] = &  -\intxt\frac{|\Delta(\bm x,\tau)|^2}{V_0}-\hbar\tr\left[\log\left(-\bm G_{\rm BCS}^{-1}\right)\right] \\
    &   +\hbar\intxtp  J^{\dagger}(\bm x,\tau)\cdot\bm G_{\rm BCS}(\bm x,\tau;\bm x',\tau';\Delta) \cdot J(\bm x',\tau') \;,
\end{split}
\end{align}
where the trace implies a summation over the Nambu space
indices as well as an integral over position and imaginary
time. The second term in the action contains all orders in
$|\Delta|^2$ and as a result the theory is thus still very
complex and impossible to solve completely. In BCS theory, we
make a saddle-point approximation and replace the pairing field
by its expectation value. In other words, we write $\Delta =
\Delta_0 +
\delta\Delta$, with $\Delta_0$ the expectation value
$\langle\Delta\rangle$ and $\delta\Delta$ representing the
fluctuations, and subsequently neglect these fluctuations. The
actual value of the BCS gap $\Delta_0$ can then be determined
by the gap equation in \eqref{eq:TTI:GapEquation}, which is
equivalent to $\delta
S^{\text{eff}}[\Delta^*,\Delta;0,0]/\delta\Delta^*|_{\Delta=\Delta_0}
= 0$, and is to be solved selfconsistently. This procedure is
of course only valid when the interaction strength is
sufficiently small.

\subsection{Fluctuations}

But what happens when the interaction strength is not small, as
is the case at unitarity? In that case we cannot neglect the
fluctuations. To deal with that situation we use in
Sec.~\ref{sec:TTI:SFselfenergy} an approach inspired by
Landau-Ginzburg theory, in which we try to find an accurate
self-energy matrix for the fermions that effectively takes all
fluctuation effects into account. In particular we need two
self-energies that contribute to the diagonal part of the exact
inverse Green's function matrix $\bm G^{-1}$, because otherwise
the normal state would correspond to an ideal Fermi mixture,
which at unitarity is not an accurate starting point for a
discussion of the instability towards superfluidity. However,
the effective interaction between the two atomic species is not
the same in the normal and superfluid states of the gas.
Therefore, also this diagonal part of the self-energy must
sufficiently deep in the superfluid state depend on the
expectation value of the pairing field or gap $\Delta_0$ and it
is important to understand how this dependence precisely comes
about. In this section we show that in principle all
interaction effects can indeed be included in a self-energy
matrix, and that also the diagonal part of this self-energy
depends explicitly on the gap. A nice and insightful way to
achieve this is by considering the appropriate Feynman
diagrams.

The diagonal parts of the Green's function matrix, i.e.,
$G_{{\rm BCS};11}$ and $G_{{\rm BCS};22}$, are dressed by the
pairing field $\Delta$. This is described by the Dyson
equation. This Dyson equation follows from inverting the
relation $\bm G_{\rm BCS}^{-1} = \bm G_0^{-1} - \Sigma_{\rm
BCS}$ and can be written as
\begin{equation}
\label{eq:TTI:DysonBCS}
    \bm G_{\rm BCS} = \bm G_0 + \bm G_0\cdot\Sigma_{\rm BCS}\cdot\bm G_{\rm BCS}\;.
\end{equation}
Diagrammatically the diagonal part of this equation can be
represented in the following way,
\begin{align}\begin{split}
    \includegraphics[scale=\fds]{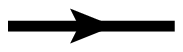} &= \includegraphics[scale=\fds]{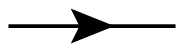} + \includegraphics[scale=\fds]{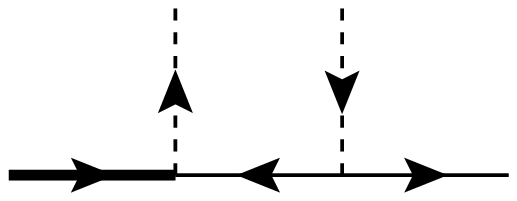} \\
    &=\includegraphics[scale=\fds]{Dyson_BCS_thin_00} + \includegraphics[scale=\fds]{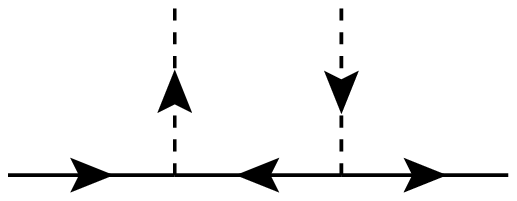} + \includegraphics[scale=\fds]{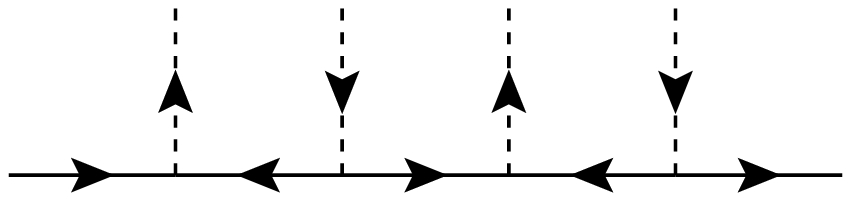} + \ldots
\end{split}\label{eq:TTI:propagatorDyson}\end{align}
Here the dashed line represents the pairing field $\Delta$ and
the direction of the arrow depicts the difference between
$\Delta$ and $\Delta^*$. The solid line represents the
noninteracting fermionic propagators $G_{0;\sigma}$, where in
this case the direction of the arrow depicts alternatingly the
propagator of the two different fermion species. The first line
of the equation shows the recurrence relation for the full
diagonal propagator and the second line shows the first three
elements originating from this Dyson equation by iteration.

In the superfluid state, the pairing field $\Delta$ has a
nonzero expectation value $\Delta_0$. In a mean-field
approximation we neglect the fluctuations and replace $\Delta$
by its expectation value $\Delta_0$. In this approximation the
diagonal propagators reduces to the standard form known from
BCS theory. However, when we take fluctuations into account we
also get self-energy corrections on the noninteracting fermion
propagators in the Dyson equation. This follows directly from
the definition of the exact fermionic propagators,
\begin{align}
\begin{split}
    \bm G_{11}(\bm x,\tau;\bm x',\tau') &=  - \langle\fld{\phi}{+}(\bm x,\tau)\flds{\phi}{+}(\bm x',\tau')\rangle\\
    &= \frac{1}{Z[0,0]} \frac{\delta}{\delta \flds{J}{+}(\bm x,\tau)}\frac{\delta}{\delta\fld{J}{+}(\bm x',\tau')}Z[J^*,J]\bigg|_{J^*=J=0}\\
    &= \frac{1}{Z[0,0]} \int\!\dd[\Delta^*]\dd[\Delta]\bm G_{{\rm BCS};11}(\bm x,\tau;\bm
    x',\tau';\Delta)\ee^{-\frac{1}{\hbar}S^{\text{eff}}[\Delta^*,\Delta;0,0]}\;,
\end{split}
\end{align}
and similarly for $\bm G_{22}$. In BCS mean-field theory we
thus have $\bm G_{11}(\bm x,\tau,\bm x',\tau')=\bm G_{{\rm
BCS};11}(\bm x,\tau,\bm x',\tau';\Delta_0)$, but at unitarity
we still have to perform a functional integral over the pairing
field to obtain the exact results.

We can represent this functional integral over the fluctuations
diagrammatically by connecting some of the $\Delta$ fields with
the pair propagator, which is determined by the effective
action $S^{\text{eff}}[\Delta^*,\Delta;0,0]$, put the other
fields equal to the expectation value $\Delta_0$, and then sum
over all possible diagrams. Because of the U(1) symmetry of the
effective action, we can only draw a pair propagator between a
$\Delta$ and a $\Delta^*$, as suggested by the arrows. The
fully dressed diagonal propagators now become,
\begin{align}
\begin{split}\label{eq:TTI:propagatorFluc}
    \includegraphics[scale=\fds]{Dyson_BCS_thick_00} &= \includegraphics[scale=\fds]{Dyson_BCS_thin_00} + \includegraphics[scale=\fds]{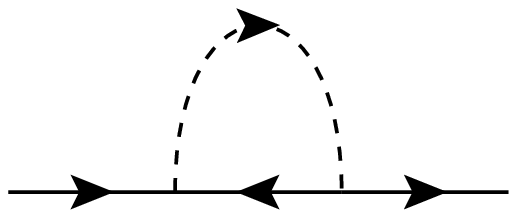} + \includegraphics[scale=\fds]{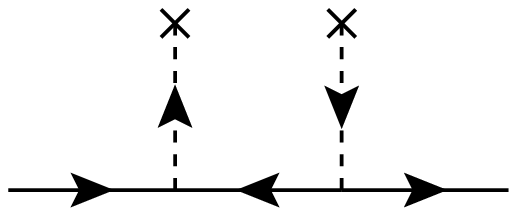} \\
        &\quad + \includegraphics[scale=\fds]{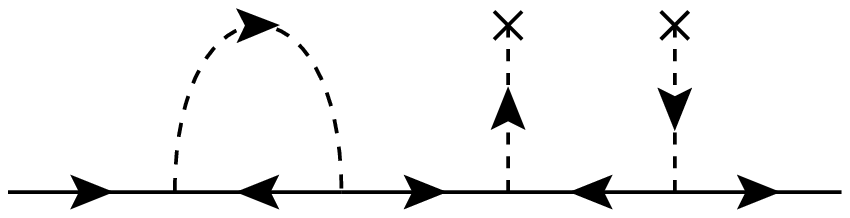} + \includegraphics[scale=\fds]{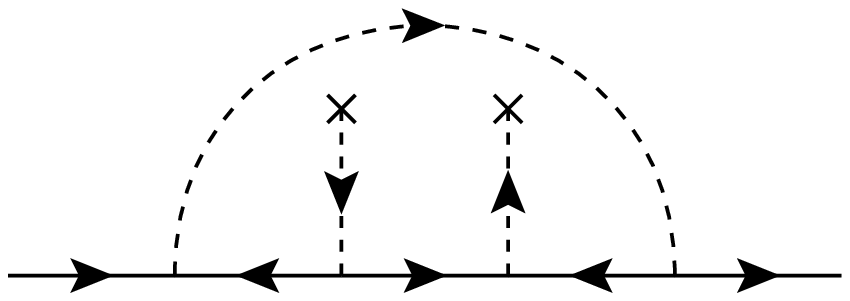} \\
        &\quad + \parbox[c][5.7em][t]{14.2em}{\includegraphics[scale=\fds]{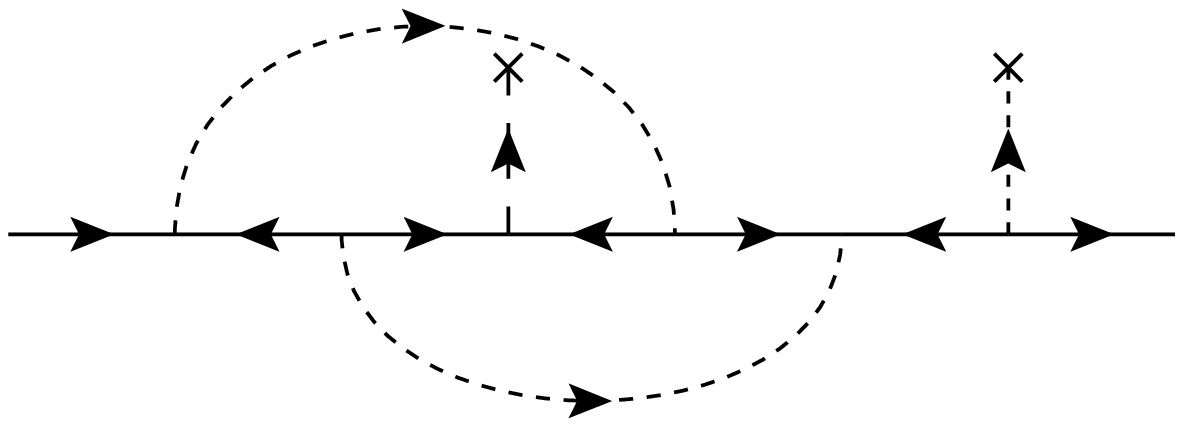}} + \ldots
\end{split}
\end{align}
Here the connected dashed lines represent the pair propagator
and the cross represents the expectation value. This series can
be resummed such that we get the exact Dyson equation
\begin{equation}
    \bm G = \bm G_0 + \bm G_0\cdot\Sigma\cdot\bm G\;,
\end{equation}
but now with an exact $2\times 2$ self-energy matrix $\Sigma$,
which contains both diagonal (normal) and off-diagonal
(anomalous) elements. For instance, the second and fifth terms
drawn in the right-hand side of \eqref{eq:TTI:propagatorFluc}
contribute to the diagonal self-energy, whereas the last term
leads to an additional contribution to the off-diagonal
self-energy. These terms thus renormalize the BCS self-energy
that is obtained from \eqref{eq:TTI:BCSselfEnergy} by replacing
$\Delta$ by $\Delta_0$. From the expectation values of the gap
inside the loops in Eq.~(\ref{eq:TTI:propagatorFluc}), we
explicitly see that the normal self-energies can be written as
a series expansion in $|\Delta_0|^2$. The same is in fact also
true for the first diagram in the right-hand side of
\eqref{eq:TTI:propagatorFluc}, because the nonlinearities in
the effective action make sure that the pair propagator already
contains all orders of $|\Delta_0|^2$. These nonlinearities
also lead to more complicated Feynman diagrams containing
higher-order (connected) correlation functions of the pair
field that are not shown here, but this does not affect our
main conclusions.

We just showed that fluctuation effects of the pair field can
be incorporated in an effective self-energy. The same
discussion can be carried out for the gap equation. This can
also be very nicely illustrated diagrammatically. The gap
equation in \eqref{eq:TTI:GapEquation} is an equation between
the expectation value of the gap and the off-diagonal or
anomalous propagator. We can again use the Dyson equation in
\eqref{eq:TTI:DysonBCS} for the anomalous propagator to study
the effects of the fluctuations on the gap equation,
\begin{equation}\label{eq:TTI:AnomalousDyson}
\begin{tabular}[t]{lcrcrcrc}
    \includegraphics[scale=\fds]{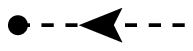} & = & \parbox[c]{4.5em}{\includegraphics[scale=\fds]{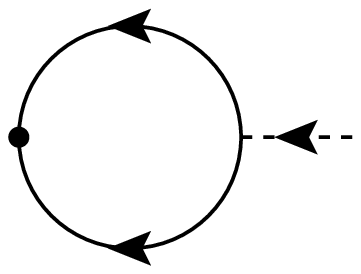}} & $+$ &\parbox[c]{4.5em}{\includegraphics[scale=\fds]{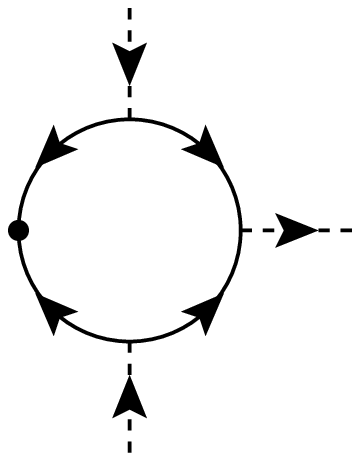}} & $+$ & \parbox[c]{4.5em}{\includegraphics[scale=\fds]{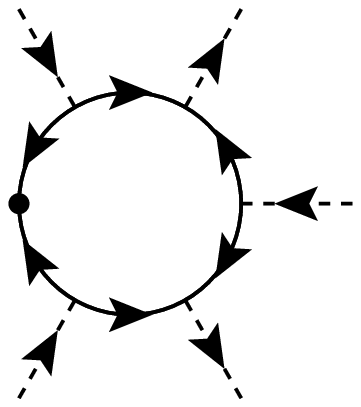}} & $+$ \ \ldots
\end{tabular}
\end{equation}
Here the small dot on the left of all diagrams represents the
fact that the gap only depends on one space-time point, i.e.,
$\langle
\Delta(\bm x,\tau)\rangle = V_0 G_{12}(\bm x,\tau;\bm x,\tau)$ due
to the point-like and instantaneous nature of the attractive
interaction.

The fluctuation effects follow again from performing the
functional integral over the $\Delta$ field, since from
\eqref{eq:TTI:GapEquation} we have that
\begin{align}
\begin{split} \langle \Delta(\bm x,\tau) \rangle
    &= \frac{V_0}{Z[0,0]}\frac{\delta}{\delta \flds{J}{-}(\bm x,\tau)}\frac{\delta}{\delta\flds{J}{+}(\bm x,\tau)}Z[J^*,J]\bigg|_{J^*=J=0}\\
    &= \frac{V_0}{Z[0,0]}\int\!\dd[\Delta^*]\dd[\Delta]\bm G_{{\rm BCS};12}(\bm x,\tau;\bm
    x,\tau;\Delta)\ee^{-\frac{1}{\hbar}S^{\text{eff}}[\Delta^*,\Delta,0,0]}\;.
\end{split}
\end{align}
The diagrammatic representation of this equation follows from
connecting some of the pair lines in
\eqref{eq:TTI:AnomalousDyson}. Again also higher-order
correlation functions of the pair field contribute, but for
simplicity we do not consider these as they do not change our
results. When we carry out this procedure we obtain
\begin{equation}
\begin{tabular}[t]{lcrcrcrcrc}
    \includegraphics[scale=\fds]{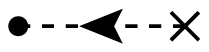} & = & \parbox[c]{4.5em}{\includegraphics[scale=\fds]{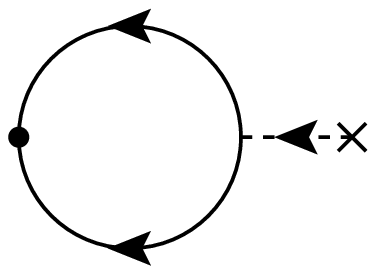}} & $+$ &\parbox[c]{4.5em}{\includegraphics[scale=\fds]{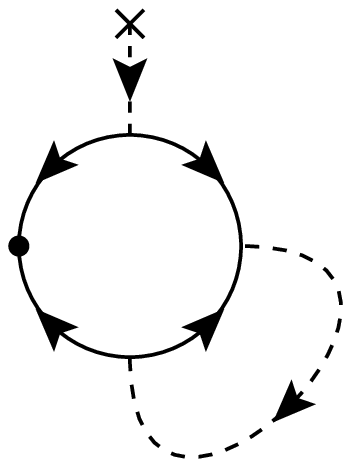}} & $+$ & \parbox[c]{4.5em}{\includegraphics[scale=\fds]{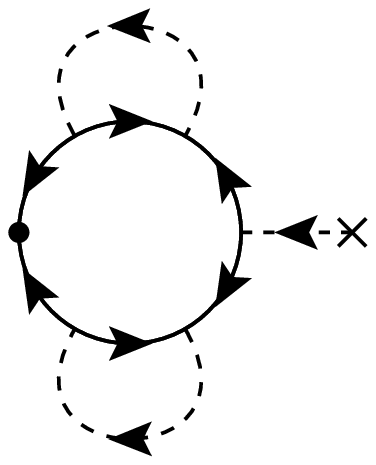}} & $+$ & \parbox[c][4.8em][t]{5.5em}{\includegraphics[scale=\fds]{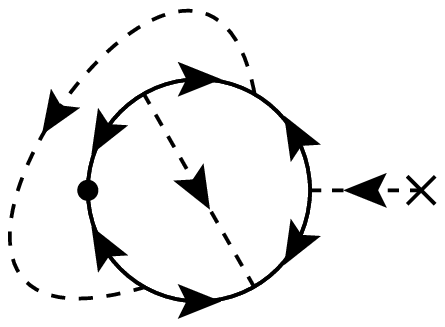}} & $+$ \ \ldots
\end{tabular}
\end{equation}
Notice that all terms are now proportional to $\Delta_0$
instead of $|\Delta_0|^2$. The first three terms in the
right-hand side can again be incorporated in a fully dressed
fermion propagator by resumming this series. The last term,
which for the gap equation behaves as a vertex correction, is
then again incorporated into the anomalous self-energy.

In the unitarity limit, these vertex corrections are important
to find the correct gap equation and, therefore, the
expectation value for the gap. Also the diagonal part of the
self-energy is important for a determination of the energy and
the densities of the Fermi mixture. There is, however, no
clear-cut way do derive these full self-energies from first
principles for the unitarity case. In this chapter, we
therefore use a more top-down approach. We will use the fact
that these self-energies exist and can be expanded in powers of
$|\Delta_0|^2$. Moreover, our previous renormalization group
theory \cite{gubbels2008rgt} has shown that for thermodynamic
quantities the self-energies can in a good approximation be
considered to be momentum and frequency independent. Combining
these observations we are ultimately able to derive an accurate
approximation to the thermodynamic potential $\Omega$ of the
unitarity Fermi mixture.

\section{The Thermodynamic Potential}

In the previous section we showed that interaction effects in
the unitary Fermi gas can be described by including appropriate
normal and anomalous self-energies into the theory. We also
discussed that this, in principle well-known fact, can be
understood as an effect of pair fluctuations. As a result the
self-energies, and in particular the normal self-energies,
depend on the gap $\Delta_0$. In addition, we showed also that
the gap equation contains vertex corrections, which cannot be
incorporated by dressing the diagonal propagators alone. This
is one important reason for deriving also the gap equation from
the thermodynamic potential, because the minimization condition
then automatically generates the correct vertex corrections.
For our purposes it is therefore crucial to realize that in
principle there exists an exact thermodynamic potential that
describes the full thermodynamics of the unitarity-limited
Fermi gas. It is, however, impossible to derive this from first
principles for this strongly interacting system, and we
therefore have to find an appropriate approximation. In this
section we will show how to arrive at such an accurate
approximation to the exact thermodynamic potential.

\subsection{Normal State}

\begin{figure}[t]
\centering
\includegraphics[width=7.5cm]{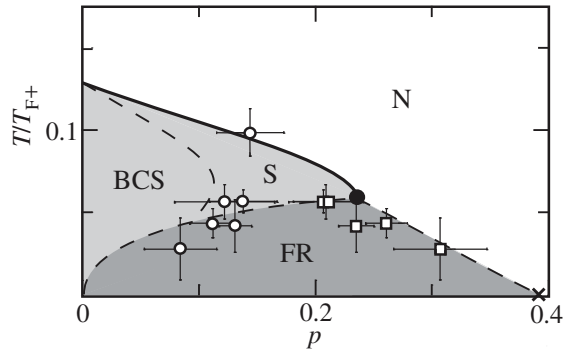}
\caption{%
The phase diagram of the unitary Fermi mixture in the
temperature-polarization plane. The Fermi temperature of the
majority species is denoted by $T_{F+}$ and the polarization $P$
equals $(N_+-N_-)/(N_++N_-)$ with $N_{\sigma}$ the number of atoms
in hyperfine state $|\sigma\rangle$. The phase diagram consists of
the normal phase (N), a forbidden region (FR) where phase
separation takes place, and the superfluid phase in which a
crossover occurs between the gapless Sarma phase (S) and the
gapped BCS phase. The solid line depicts the line of second-order
phase transitions \cite{gubbels2008rgt}, the dashed line gives the
boundary of the forbidden region associated with the first-order
phase transitions, and the black dot represents the tricritical
point. The open squares and circles are experimental data points
\cite{shin2008pdt}.} \label{fig:TTI:phaseDiagram}
\end{figure}

Despite the strong interaction, it is now rather well
established that BCS mean-field theory gives the correct
qualitative description of the unitarity limit, at least at the
temperatures accessible to the state-of-the-art experiments.
Therefore a reasonable starting point for the approximation of
the thermodynamic potential is this mean-field theory. From
experiments, renormalization group theory, and several
Monte-Carlo calculations it is found that the phase diagram has
the following features, as illustrated in
Fig.~\ref{fig:TTI:phaseDiagram}. At zero temperature both
experiments and theoretical calculations find a first-order
phase transition at a local critical polarization $P_{\text{c}}
\simeq 0.4$. In the balanced situation $P=0$ both find a
second-order transition at a critical temperature of about
$T_{\text{c}} \simeq 0.15 T_{\text{F}}$ \cite{burovski2006ctt}.
These second and first-order transition lines should then be
connected by a tricritical point, which is confirmed in
experiments and by renormalization group theory.

The thermodynamic potential in BCS theory leads to exactly the
same qualitative behavior of the phase diagram, although the
critical temperatures and critical polarizations are off by
almost an order of magnitude and would not be visible in the
window shown in Fig.~\ref{fig:TTI:phaseDiagram}. We therefore
start with BCS theory, after which we systematically include
the dominant interaction effects that are still missing. At
unitarity the BCS energy functional is
\begin{align}
\begin{split}\label{eq:TTI:BCSfreeEnergy}
    \Omega_{\text{\rm BCS}}[\Delta; \mu, h] =&
    \sum_{\bm k}\left(\epsilon_{\bm k}-\mu-\hbar\omega_{\bm k}
                  +\frac{|\Delta|^2}{2\epsilon_{\bm k}}\right) \\
    &-k_{\text{B}}T\sum_{\sigma, \bm k}
      \log\left(1+\ee^{-\hbar\omega_{\bm k, \sigma}/k_{\text
      B}T}\right)~,
\end{split}
\end{align}
where $\epsilon_{\bm k} = \hbar^2\bm k^2/2m$, $m$ is the atomic
mass, and the superfluid dispersion is given by the well-known
BCS formula, $\hbar\omega_{\bm k}=\sqrt{(\epsilon_{\bm
k}-\mu)^2+|\Delta|^2}$. The second term in the right-hand side
contains also a sum over the pseudospin projection
$\sigma=\pm$, and represents the contribution due to an ideal
gas of quasiparticles with the quasiparticle dispersion of the
two spin states given by $\hbar\omega_{\bm k,\sigma} =
\hbar\omega_{\bm k} - \sigma h$. Finally, we introduced the
average chemical potential $\mu=(\mu_++\mu_-)/2$ and half the
chemical potential difference $h=(\mu_+-\mu_-)/2$ that acts as
an effective magnetic field on the pseudospin as the
quasiparticle dispersion $\hbar\omega_{\bm k,\sigma}$ clearly
shows.

In BCS theory, the normal state is treated as an ideal Fermi
gas, thus no interactions are taken into account. This is not
correct in the unitarity limit. As discussed above, these
interaction effects can be described by two self-energies. The
imbalanced normal phase in the unitary limit, has been studied
with Monte Carlo methods \cite{lobo2006nsp}. From this, the
equation of state can be determined. If we can find the
self-energies such that it reproduces the same equation of
state for the theory, we have effectively taken all interaction
effect in the normal phase into account.

For momentum and frequency independent self-energies, the
self-energies can be incorporated in the theory of an ideal
Fermi gas, by just changing the chemical potential. We thus
replace the chemical potentials as
\begin{align}
   \mu_{\sigma}' = \mu_{\sigma} - \hbar\Sigma_{\sigma}\;.
\end{align}
Here $\mu_{\sigma}'$ is the effective chemical potential and
$\Sigma_{\sigma}$ the self-energy for species $\sigma$.
Inspired by Hartree-Fock theory we would write down an {\it
ansatz} for the self-energy of species $\sigma$ that is
proportional to the density of species $-\sigma$
\cite{gubbels2006spt}. However, the densities are in a
grand-canonical setting calculated by taking the derivative of
$\Omega$ with respect to the chemical potentials, i.e.,
$N_\sigma = - \partial \Omega/
\partial\mu_{\sigma}$. It is therefore preferable to write the
self-energies as a function of the chemical potentials only. By
considering terms with the correct units that incorporate the
Hartree-Fock-like feature mentioned above, we find that the
following self-energies gives rise to the correct equation of
state of the strongly interacting normal phase,
\begin{align}\label{eq:TTI:LDA_SelfEnergy}
    \mu_{\sigma}'=\mu_{\sigma} +
    \frac{3}{5}A\frac{(\mu_{-\sigma}')^{2}}{\mu_+'+\mu_-'}\;.
\end{align}
The prefactor can be determined from the self-energy of a
single minority atom in the presence of a Fermi sea of majority
atoms and equals $A \simeq 0.96$
\cite{lobo2006nsp,chevy2006upd,combescot2007nsh,gubbels2008rgt}.
Explicitly in terms of $\mu$ and $h$, these relations imply
that
\begin{align}
\label{eq:TTI:LDA_SelfEnergy2}
\begin{split}
    \mu' & = \mu \left(1 - \frac{5-3A}{10-3A}
           + \frac{5 \sqrt{(5+3A)^2+3A(10-3A){(h/\mu)^{2^{\phantom 2}}}}}{(10-3A)(5+3A)}\right) \;,\\
    h'   & = h \left(1 - \frac{3A}{5+3A}\right)\;.
\end{split}
\end{align}

\begin{figure}[t]
    \centering
    \includegraphics[width=7.5cm]{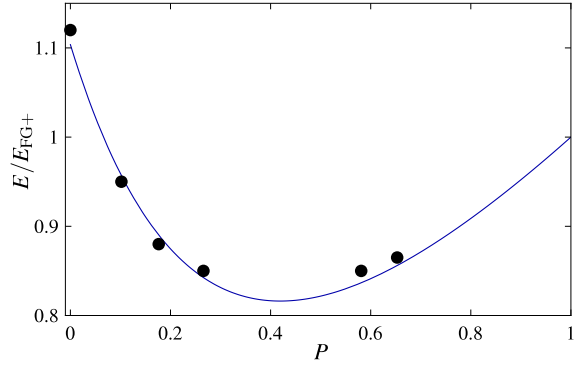}
    \caption{
    The equation of state of the normal phase at zero temperature,
    with on the horizontal axis
    the polarization and on the vertical axis the energy.
    The dots are the Monte-Carlo data from Ref. \cite{lobo2006nsp} and the
    line is the equation of state found with the use of the effective chemical
    potentials as defined in \eqref{eq:TTI:LDA_SelfEnergy2}.
    The energy is given by $E = \Omega +\mu_+N_++\mu_-N_-$ and
    is scaled with the ideal gas energy of the majority component of the mixture
    $E_{{\rm FG}+}=\frac{3}{5}E_{{\rm F}+}N_+$ and $E_{{\rm F}+}$
    the Fermi energy of the majority species.
    }
    \label{fig:TTI:EOS_MC}
\end{figure}

In Fig.~\ref{fig:TTI:EOS_MC} the resulting energy of the
mixture determined from the thermodynamic potential
$\Omega(\mu,h,T,V)=\Omega_{\text{\rm BCS}}[0; \mu', h']$ at
zero temperature is plotted as a function of the polarization.
This figure shows the excellent agreement between the
Monte-Carlo data and the {\it ansatz} from
\eqref{eq:TTI:LDA_SelfEnergy2}. In the next section we discuss
how these self-energies can be further improved when we also
consider the effects of pairing in the superfluid state.

\subsection{Superfluid State}\label{sec:TTI:SFselfenergy}

When the temperature is low enough and the imbalance not too
large, the unitary Fermi gas becomes superfluid. In the
unitarity limit, the scattering length goes to infinity and is
no longer a relevant length scale. In fact, in the homogeneous
situation, the (average) Fermi energy is the only energy scale
in the problem. This makes the system universal and as a
result, we can write most thermodynamic properties of the
system in terms of this Fermi energy \cite{chevy2006upd}.

In Sec.~\ref{sec:TTI:LG} we showed that the self-energies can
be explicitly written as a power series in $|\Delta|^2$. The
straightforward first step to incorporate these superfluid gap
corrections to the self-energy is to take the first term in
$|\Delta|^2$ into account \cite{bulgac2008ufs}. We subtract
this from the effective chemical potential in
\eqref{eq:TTI:LDA_SelfEnergy2} as
\begin{align}
    \mu'(\mu,h,\Delta) = \mu'(\mu,h,0) - B\frac{|\Delta|^2}{\mu'(\mu,h,0)}
\end{align}
and $B$ a constant to be determined next. For this we use one
simple but important piece of information, namely the value of
the thermodynamic potential in the balanced superfluid minimum.
From experiments and Monte-Carlo calculations this minimum is
known to be
\begin{align}\label{eq:TTI:SF_energyfunc_minimum}
    \Omega = -\frac{4\sqrt{2} \mu
    ^{5/2}m^{3/2}}{15 \pi ^2 \hbar^3(1+\beta)^{3/2}} V \equiv \Omega_{\text{cr}}\;,
\end{align}
with $V$ the volume and $\beta \simeq -0.58$ a universal
number. Matching the energy in the minimum is important,
because this ensures a correct energy balance between the
(imbalanced) normal state and the superfluid state and
therefore the correct location of the first-order phase
transition at low temperature. From experiments and several
theoretical calculations, it is now believed that at low
temperatures the superfluid state is balanced. Thus, to find
the transition we should compare the energy in the balanced
superfluid with the normal state energy, for which we have
already a description that agrees with the Monte-Carlo equation
of state and thus has the correct energy. This condition fixes
the unknown constant to $B \simeq 0.21$, which follows directly
from the zero-temperature minimum of $\Omega_{\text{\rm
BCS}}[\Delta; \mu', 0]$ in \eqref{eq:TTI:BCSfreeEnergy} with
both self-energy corrections subtracted from the chemical
potential.

At this point our construction, where everything is explicitly
written in terms of the chemical potentials $\mu$ and $h$,
gives rise to a problem: The superfluid in the minimum of the
thermodynamic potential turns out not to be balanced at low
temperatures for $h \neq 0$. This problem originates from the
normal self-energies in \eqref{eq:TTI:LDA_SelfEnergy2} which
explicitly depends on the chemical potential difference $h$. It
is in particular the renormalization of the average chemical
potential which depends on $h$, thus $\mu'(\mu, h,\Delta)$.
This problem could have been avoided by making an {\it ansatz}
in terms of the densities instead of the chemical potentials,
which would automatically have resulted in a balanced
superfluid \cite{jeroen2009ifm}. This follows directly from the
fact that BCS theory already gives a balanced superfluid at low
temperature and the dependence on imbalance in the
self-energies is thus suppressed in the superfluid state.
Physically, the problem is that the formation of a
Bose-Einstein condensate of Cooper pairs gives the superfluid
state a strong preference for equal densities of the two spin
states, which is not present in the normal state. To
incorporate this extra piece of physics into the theory, we
need to add an extra $|\Delta|^2$ dependence to the model to
ensure a balanced superfluid in the minimum of the
thermodynamic potential. There are several ways to achieve
this, but an exponential suppression of the $h$ dependence in
$\mu'$ turns out to give the best interpolation between the
various known regimes. Technically this is achieved, by
replacing $h$ in $\mu'(\mu,h,\Delta)$ by
$h\exp{(-|4\Delta|^2/\mu^2)}$. The factor of $4$ in the
exponent is somewhat arbitrary, but should be large enough to
make the $h$ dependence in the ground-state superfluid minimum
negligible.

\begin{figure}[t]
    \sidecaption[t]
    \centering
    \includegraphics[width=7.4cm]{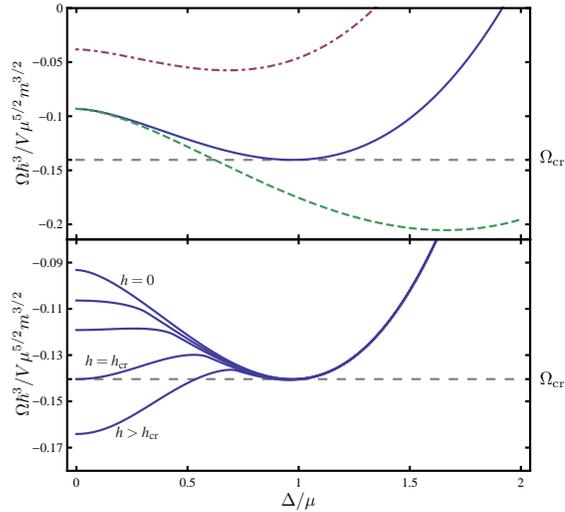}
    \caption{
    The zero-temperature thermodynamical potential functional as a function of the order
    parameter $\Delta$. The upper panel illustrates the balanced case,
    where the dash-dotted line is the usual BCS result, the dashed
    line incorporates only the normal-state self-energy effects, and
    the solid line includes also the superfluid
    self-energy correction. In the lower panel the
    energy functional is shown for various values of the chemical
    potential difference $h$, with $h_{\text{cr}} \simeq 0.94\mu$ its
    critical value.}\label{fig:TTI:energyFunctionals}
\end{figure}

We now have included the self-energy effects in both the normal
state as well as in the superfluid state. This results in an
approximation for the thermodynamic potential which has the
correct equation of state in the normal phase, the correct
energy minimum for the superfluid phase, and interpolates
between these two in a manner that incorporates all the known
physical properties of the system. In
Fig.~\ref{fig:TTI:energyFunctionals} the resulting
thermodynamic potential is plotted for several values of $h$
and at zero temperature. As a check we can compute the critical
polarization which gives about $P \simeq 0.4$ as desired. Also
the universal number $\zeta=\Delta_0/\mu$ of the balanced
superfluid ground state has a very reasonable value. Here we
find $0.97$ while Monte Carlo gives $1.07\pm0.15$
\cite{carlson2008spg,carlson2005atf}. In principle, we can
easily correct for this difference by including a small
correction to the anomalous self-energy, but in view of the
already rather good agreement with the Monte-Carlo results we
refrain from doing so in the following.

\begin{figure}[t]
    \centering
    \includegraphics[width=8cm]{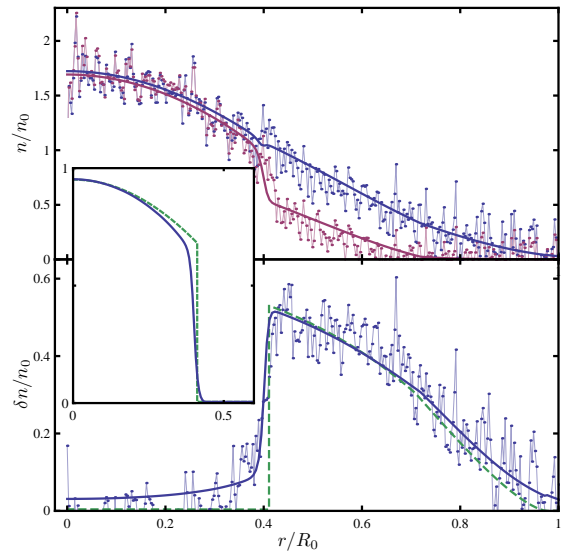}
    \caption{
    (Color online) The density profiles
    of a unitary mixture with polarization $P \simeq 0.44$ in a harmonic trap.
    The upper figure shows the majority and minority densities as a
    function of the position in the trap. The lower figure shows the
    density difference, where the theoretical curves show the results
    both within the LDA (dashed line) and for our theory (solid line) that
    goes beyond this approximation and, therefore, allows for a
    substantial better agreement with experiment. The inset shows the
    BCS gap parameter $\Delta_0(r)/\Delta_0(0)$ both for the LDA (dashed line) and our
    theory (solid line). The experimental data points and
    scaling are from Shin {\it et al}.\ \cite{shin2008pdt}.
    }\label{fig:TTI:profiles}
\end{figure}

A large region of the trapped unitary Fermi gas can be well
described using the local-density approximation. However, near
the interface of a first-order phase transition, this
approximation always breaks down, as it leads to an unphysical
discontinuity in the density profiles. The thermodynamic
potential we constructed so far also describes the system out
of equilibrium, i.e., with $\Delta$ not in a minimum of the
thermodynamic potential, which is precisely what happens near
the interface. But in order to describe the interface properly,
we need to go beyond the LDA by including also a gradient term
for $\Delta$ in the thermodynamic potential,
\begin{align}\label{eq:TTI:gradientTerm}
    \Omega[\Delta;\mu,h] =& \int \dd \bm{x}~
    \left(\frac{1}{2} \gamma(\mu,h)|\nabla\Delta(\bm{x})|^2
     + \omega_{\text{\rm BCS}}[\Delta(\bm{x});\mu',h'] \right) \;,
\end{align}
where $\omega_{\text{BCS}}$ denoted the homogeneous
thermodynamic potential density $\Omega_{\text{BCS}}/V$ and
$\hbar\gamma(\mu,h)\sqrt{\mu/m}$ is a positive function of the
ratio $h/\mu$ only, due to the universal nature of the Fermi
mixture at unitarity. The functional minimum of this new
thermodynamic potential gives a smooth transition at the
interface, instead of the discontinuous step obtained within
the LDA. A careful inspection of the interface in the data of
Shin {\it et al}.\ \cite{shin2008pdt}, cf.\
Fig.~\ref{fig:TTI:profiles}, also reveals that the interface is
not a sharp step. This is most clear in the data for the
density difference, since the noise in the density difference
is much smaller than in the total density. This has to do with
the experimental procedure used, which only measures the
density difference directly. Such a smooth transition arises
also in the self-consistent Bogoliubov-de Gennes equations. But
these lead then also to oscillations in the order parameter and
the densities, due to the proximity effect
\cite{mcmillan1968tsn}. This is not observed experimentally.
Oscillations will also occur in our Landau-Ginzburg approach if
$\gamma(\mu,h)<0$. However, we have checked both with the above
theory as well as with renormalization group calculations
\cite{gubbels2008rgt} that $\gamma(\mu,h)$ is positive. This
agrees with the phase diagram of the imbalanced Fermi mixture
containing a tricritical point and not a Lifshitz point in the
unitarity limit \cite{jildou2009lpp}.

We restrict ourselves here to a gradient term that is of second
order in $\Delta$ and also of second order in the gradients.
There are of course higher-order gradient terms that may
contribute quantitatively \cite{stoof1993tgl}, but the
leading-order physics is captured in this way due to the
absence of a Lifshitz transition. One way to compute the
coefficient $\gamma(\mu,h)$ is to use the fact that in
equilibrium this coefficient can be exactly related to the
superfluid stiffness, and therefore the superfluid mass density
$\rho_{\text{s}}$, by $\gamma =\hbar^2
\rho_{\text{s}}/4m^2|\langle\Delta\rangle|^2$. At zero temperature
it gives the simple result that $\gamma(\mu,h)=
\sqrt{m/2\mu}/6\pi^2\hbar\zeta^2(1+\beta)^{3/2}$, with $\beta$ and
$\zeta$ universal constants as defined earlier. With this
result for $\gamma$ our thermodynamic potential functional in
\eqref{eq:TTI:gradientTerm} contains no longer any free
parameters and can now be confronted with experiments. The
result of this comparison, at a realistic temperature of about
one third the tricritical temperature $T_{c3}$, is shown in
Fig.~\ref{fig:TTI:profiles} and turns out to be excellent.

\section{Applications}

We have thus constructed an accurate approximation to the exact
thermodynamic potential of the imbalanced Fermi mixture at
unitarity. With a simple {\it ansatz} for the self-energies we
can describe both the homogeneous normal and superfluid phase
at zero and nonzero temperatures. Moreover, the description is
also valid out of equilibrium, i.e., when the value of the gap
is not in a minimum of the thermodynamic potential. By
including also the energy cost for gradients of the gap
parameter we have a Landau-Ginzburg-like theory that can
describe the inhomogeneous situation that is used in
experiments \cite{hulet2006pps, shin2008pdt} in a manner that
goes beyond the local-density approximation.

In this section we use the thermodynamic potential
$\Omega[\Delta;\mu,h]$ from \eqref{eq:TTI:gradientTerm} to
investigate the properties of the superfluid-normal interface.
First, we consider the trap to be spherically symmetric and in
that case calculate the surface tension of the interface. This
is an important quantity that has been put forward
\cite{silva2006stu,hulet2006pps} as a possible explanation for
the deformations of the superfluid core observed by Partridge
{\it et al}.\ \cite{hulet2006pps}. Second, we then show how the
anisotropy of the trap can be incorporated and study the effect
of this anisotropy on the equilibrium gap profile
$\Delta_0({\bf x})$. In this section we for simplicity always
take the gap $\Delta_0({\bf x})$ to be real, which does not
lead to any loss of generality for the applications that we
consider here.

\subsection{Interface and Surface Tension}

The fact that we are able to study the superfluid-normal
interface beyond the LDA, makes it possible for us to also
determine the surface tension. The surface tension is
determined by the difference in thermodynamic potential between
a one-dimensional LDA result with a discontinuous step in
$\Delta_0({\bf x})$ and our Landau-Ginzburg theory with a
smooth profile for the order parameter $\Delta_0({\bf x})$. In
actual experiments, however, the width of the interface is
rather small compared to the size of the whole atomic cloud.
This makes it possible to compute the surface tension by
considering a flat interface in a homogeneous system rather
than a curved interface in the trap. In the homogeneous case,
such an interface occurs only when the imbalance is critical,
i.e., when $h=h_{\rm cr}(\mu)=\kappa\mu$ with $\kappa$ another
universal number, for which we have obtained $\kappa \simeq
0.94$. This means that the thermodynamic potential of the
normal state minimum is exactly equal to the thermodynamic
potential of the superfluid state minimum. The surface tension
is then the difference in thermodynamic potential between a
system that stays in one minimum and one that goes near the
interface from one minimum to the other.

How the system achieves the latter is determined by minimizing
the thermodynamic potential,
\begin{align}\label{eq:TTI:gapEOM}
   \left. \frac{\delta\Omega[\Delta;\mu,h_{\rm cr}]}{\delta\Delta(z)}\right|_{\Delta=\Delta_0}
    = \frac{\partial\omega_{\text{BCS}}[\Delta_0(z);\mu',h']}{\partial\Delta}
     - \gamma(\mu,h_{\rm cr}) \frac{\partial^2}{\partial z^2}\Delta_0(z) =
     0\;.
\end{align}
In principle, this highly nonlinear equation can be numerically
solved, to get a hyperbolic tangent-like function for
$\Delta_0(z)$ that on the normal side of the interface
approaches zero and on the superfluid side approaches the
equilibrium position of the superfluid minimum that we simply
denote by $\Delta_0$. Fortunately, however, this solution is
not needed to compute the surface tension, because the surface
tension can be conveniently written as
\begin{align}
    \sigma(\mu) = & \int_{-\infty}^{\infty}\!\dd z \;
     \left(\omega[\Delta_0(z);\mu,h_{\rm cr}] - \omega[0;\mu,h_{\rm cr}]\right)
     \;,
\end{align}
where $\omega=\Omega/V$ is the thermodynamic potential density.
This equation can be rewritten as an integral over $\Delta$,
knowing that $\Delta_0(z)$ is a monotonically increasing
function between zero and $\Delta_0$. Using also the first
integral of \eqref{eq:TTI:gapEOM}, we end up with
\begin{align}
    \sigma(\mu) = & \sqrt{2\gamma(\mu,h_{\rm cr})}\int_0^{\Delta_0}\!\dd
    \Delta\sqrt{\omega_{\text{BCS}}[\Delta;\mu',h'] -
    \omega_{\text{BCS}}[0;\mu',h']}\;.
\end{align}
This is clearly independent of the actual shape of the
interface. The surface tension thus only depends directly on
$\sqrt{\gamma(\mu,h_{\rm cr})}$ and on the shape of the barrier
in between the two minima of the thermodynamic potential. It is
useful to write the surface tension in a dimensionless form. We
define this as $\sigma(\mu) = \eta (m/\hbar^2)\mu^2$, with
$\eta$ a dimensionless number. This number depends only on the
temperature. In a trap, the relevant chemical potential is the
one at the position of the interface. This location is also
dependent on the polarization of the mixture and in that manner
also the surface tension will inherit in a trap a dependence on
the polarization \cite{haque2007tfc}.

\begin{figure}[t]
    \centering
    \includegraphics[width=8cm]{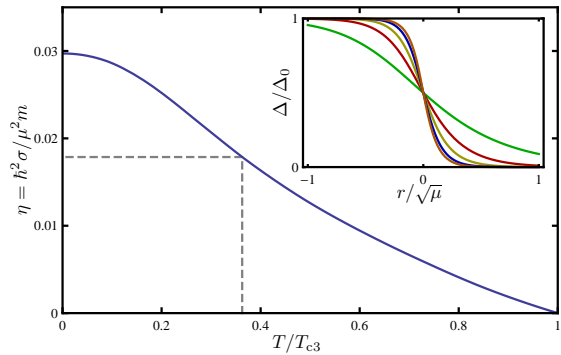}
    \caption{
    The surface tension as a function of the temperature, computed
    in the homogeneous case at unitarity. The temperature
    is scaled by the temperature of the tricritical point $T_{\text{c}3}$. The dashed line shows
    the value used to compare with experiments in Fig.~\ref{fig:TTI:profiles}. The inset shows the gap around the
    interface for several temperatures $0.9$, $0.7$, $0.5$, $0.25$ and $0.01$ $T_{\text{c}3}$,
    respectively.}\label{fig:TTI:LG_SurfaceTension}
\end{figure}

The surface tension of this model is plotted in
Fig.~\ref{fig:TTI:LG_SurfaceTension} as a function of the
temperature. Here the surface tension is plotted in its
dimensionless form. In this form it was previously found that
for the experiment of Partridge {\it et al}.\ $\eta
\simeq 0.6$ \cite{partridge2006dtf}. This was extracted from the
large deformations of the superfluid core observed in that
experiment. The experiment of Shin {\it et al}.\ does not show
any deformation, which puts an upper bound on $\eta$ of about
$0.1$ \cite{haque2007tfc,baur2009tns}. At the tricritical point
the surface tension vanishes and at zero temperature it is
about $\eta\simeq 0.03$. For a more realistic temperature of
about $0.3 T_{\text{c3}}$ we find $\eta \simeq 0.02$ which is
significantly smaller than the surface tension that would cause
a substantial deformation. This is thus in agreement with the
experiment of Shin {\it et al}.\ \cite{shin2008pdt}.

We now give a more detailed discussion of our analysis of the
density profiles observed by Shin {\it et al}. In experiments
the cloud is trapped in an anisotropic harmonic potential,
which is cigar shaped, and in the axial direction less steep
than in the radial direction. However, since the atomic cloud
shows no deformations in this case we can in a good
approximation take the trap to be spherically symmetric. The
order parameter then depends only on the radius, and the total
thermodynamic potential is given by integrating our
Landau-Ginzburg-like thermodynamic potential density over the
trap volume. To account for the trap potential in the energy
functional we let the average chemical potential depend on the
radius, such that we have $\mu(r) = \mu - V(r)$, with $V(r)$
the effectively isotropic harmonic potential.

To find the order parameter as a function of the radius we have
to minimize the energy functional with respect to the order
parameter, or
$\delta\Omega[\Delta;\mu,h]/\delta\Delta(r)|_{\Delta=\Delta_0}
= 0$. This gives a second-order differential equation for
$\Delta_0(r)$ as we have seen. Solving this Euler-Lagrange
equation, with the proper boundary conditions in the center of
the trap, gives a profile for $\Delta_0$ that is shown in the
inset of Fig.~\ref{fig:TTI:profiles}. This profile of the order
parameter is much smoother than the discontinuous step one
obtains within the LDA that is also shown in
Fig.~\ref{fig:TTI:profiles}. Besides this, there are two more
aspects that deserve some attention. First, we notice that the
value of the gap at the original LDA-interface is decreased by
almost a factor of three and, second, the gap penetrates into
the area originally seen as the normal phase. This behavior
makes the gap for a small region smaller than $h'$, giving
locally rise to a gapless superfluid, which implies a
stabilization of the Sarma phase.

Before discussing this particular physics, we focus first on
the density difference. To obtain the density profiles within
our theory, the thermodynamic relation
$n_{\sigma}(r)=\partial\omega_{\rm
BCS}/\partial\mu_{\sigma}(r)$ is used, where
$n_{\sigma}=N_{\sigma}/V$ is the density of particles in state
$|\sigma\rangle$ and $\mu_{\sigma}(r)=\mu_{\sigma} - V(r)$ the
associated local chemical potential. It is important that,
because of the self-energy effects, we cannot use the standard
BCS formulas for the density, but really have to differentiate
the thermodynamic potential. In BCS theory this would of course
be equivalent. Given the density profiles, the comparison
between theory and experiment can be made and is ultimately
shown in Fig.~\ref{fig:TTI:profiles}. Overall the agreement is
very good. Theoretically the interface appears to be somewhat
sharper than observed. This can be due to higher-order gradient
terms, that are neglected in the calculation and that would
give an additional energy penalty for a spatial variation of
the order parameter. There are also experimental effects that
could make the interface appear broader, for instance, the
spatial resolution of the tomographic reconstruction or the
accuracy of the elliptical averaging \cite{priv:ket}.

The Landau-Ginzburg-like approach presented here, shows some
new features compared to the LDA. One interesting feature is
the kink, that is visible in the majority density profile shown
in Fig.~\ref{fig:TTI:profiles}. Notice that this kink appears
\emph{before} the original (LDA) phase transition from the
superfluid to the normal phase. This kink signals a crossover
to a new exotic phase, namely the gapless Sarma phase. Note
that at zero temperature this crossover becomes a true quantum
phase transition. At the crossover, the order parameter becomes
smaller than the renormalized chemical potential difference
$h'$ and the unitarity limited attraction is no longer able to
fully overcome the frustration induced by the imbalance. As a
result the gas becomes a polarized superfluid. Because the gap
$\Delta$ is smaller than $h'$ this corresponds to a gapless
superconductor. In a homogeneous situation this can, far below
the tricritical temperature, never be a stable state as shown
in Fig.~\ref{fig:TTI:phaseDiagram}. However, because of the
inhomogeneity induced by the confinement of the gas, the gap is
at the interface forced to move away from the local minimum of
the thermodynamic potential and ultimately becomes smaller than
$h'$. The Sarma state is now locally stabilized even at these
low temperatures. Notice that this is a feature of the smooth
behavior of the gap and that the presence of the Sarma phase
thus does not depend on the quantitative details of the energy
functional $\Omega[\Delta;\mu, h]$.

\subsection{Deformation}

When the surface tension is sufficiently small or when the
aspect ratio of the external potential of the system is close
to one, the gap profile $\Delta_0({\bf x})$ will closely follow
the equipotential surfaces of the external trap and can be
reasonably well approximated by a function of a single variable
only. This can be achieved by scaling away the anisotropy of
the external potential and introducing the effective radius
\begin{equation}\label{eq:TTI:ellipticalSymmetry}
    R^2 = x^2 + y^2 + \left(\frac{z}{\alpha}\right)^2\;,
\end{equation}
with $\alpha$ the aspect ratio of the trap. However, when the
aspect ratio is large, this might not always be valid. In the
experiment of Partridge {\it et al}.\ \cite{partridge2006dtf},
an aspect ratio of about 45 is used, and dramatic deviations
between the equipotential surfaces and the shape of the
superfluid core are observed. This can be explained by a large
surface tension \cite{haque2007tfc}, but as we have just seen
the required large value of $\eta$ cannot yet be understood
from a microscopic theory. Another possibility is that the gas
has ended up in a metastable state in which the shape of the
gap parameter differs from the equipotential surfaces of the
external potential \cite{baksmaty2010cms}.

The latter possibility is something that can also be
investigated using the thermodynamic potential that we have
just derived. To do so, we first study the linear response of
the system when we also allow the gap profile to depend on more
(angular) variables then the effective radius $R$. After that
we also look at gap profiles with a different aspect ratio than
the external potential. It appears from our analysis that our
present Landau-Ginzburg-like approach gives indeed rise to
small deviations in the gap shape. However, it does not exhibit
a metastable state with a deviation that is as large as seen in
the experiment of Partridge {\it et al}.

\subsubsection{Linear Response}

The harmonic potential used in the experiments has an
elliptical symmetry, which means that it can be written as a
function of a single coordinate $R$ as defined in
\eqref{eq:TTI:ellipticalSymmetry}. As a consequence, the local
thermodynamic potential also only depends explicitly on this
$R$. Therefore, in the local-density approximation, the gap
parameter can only depend on $R$ as well. When we go beyond the
LDA, by including gradient terms in the theory, this symmetry
is explicitly broken.

In this section we first perform the above-mentioned scaling of
the axial coordinate. After that we can treat the beyond-LDA
corrections of the gap profile as perturbations on the
symmetric solution that can be expanded in the form of
spherical harmonics as
\begin{equation}\label{eq:TTI:sphericalExpansion}
    \Delta_0(\bm x) = \sum_{l} \frac{D_{l}(R)}{R} Y_{l0}(\theta,\phi)\;.
\end{equation}
Since the trap is rotationally symmetric around the z-axis, the
gap profile does not depend on the azimuthal angle $\phi$ and
we are allowed to take $m=0$ in the expansion in
\eqref{eq:TTI:sphericalExpansion}. Also the mirror symmetry in
the $x$-$y$ plane causes all coefficients with odd $l$ to be
zero. We will now assume that the elliptically symmetric part
$D_0(R)$ is much larger than the part with coefficients $l>0$
and for simplicity only look at the first anisotropic
perturbation $D_2(R)$.

To describe the deformations we have thus chosen spherical
coordinates, but with the $z$-coordinate defined as $z =
\alpha R \cos{\theta}$. This coordinate system is not orthogonal
and gives rise to a coupling between the spherical harmonics
due to the gradient terms. The Jacobian is given by $\alpha R^2
\sin{\theta}$. The gradient terms in the thermodynamic potential
can be written in these coordinates as
\begin{align}
    \Omega_{\text{gr}}[\Delta_0] & \equiv \int\!\dd\bm x  \frac{\gamma({\bf x})}{2}\left|\nabla\Delta_0({\bf x})\right|^2 \\
    & \simeq - \frac{\alpha}{2}\int_0^{\infty}\!\dd R \bigg\{\gamma_0 D_0(R) \frac{d^2}{dR^2}D_0(R) + \gamma_2 D_2(R)\left(\frac{d^2}{dR^2}D_2(R)-\frac{6}{R^2}D_2(R)\right)  \nonumber \\
    &\quad\quad\quad\quad\quad\quad +\gamma_{02}
    D_0(R)\left(2\frac{d^2}{dR^2}D_2(R)+\frac{6}{R}\frac{d}{dR}D_2(R)+\frac{3}{R^2}D_2(R)
    \right)\bigg\}\;. \nonumber
\end{align}
Here we suppressed for convenience the dependence on the
chemical potentials and approximated the stiffness $\gamma({\bf
x})$ by its value at the location of the interface, that we
from now on denote simply by $\gamma$. The latter is a very
good approximation in practice, because for the traps of
interest the width of the superfluid-normal interface is much
smaller that the typical length scale on which the trapping
potential varies. Furthermore, we defined the various different
effective stiffnesses as
\begin{align}\label{eq:TTI:gradienCoefs}
    \gamma_0 &= \left(\frac{2}{3}+\frac{1}{3}\frac{1}{\alpha^2}\right)\gamma, &
    \gamma_2 & = \left(\frac{10}{21}+\frac{11}{21}\frac{1}{\alpha^2}\right)\gamma, &
    \gamma_{02} = -\frac{2}{3\sqrt{5}}\left(1-\frac{1}{\alpha^2}\right)\gamma.
\end{align}
This can naturally be extended to general $l$, where every
$D_l$ is coupled to $D_{l+2}$, but we do not need that
extension here.

As indicated above, we want to treat $D_2$ as a small
perturbation in linear-response theory. To achieve this we need
to expand the local part of the thermodynamic potential in
terms of $D_2$. This is straightforward and is given by,
\begin{equation}
    \Omega_{\text{loc}}[\Delta_0] = \alpha\int_0^{\infty}\!\dd R\left\{ 4\pi R^2 \omega_{\rm BCS}[\Delta_0(R);R] + \frac{1}{2}\frac{\partial^2\omega_{\rm BCS}[\Delta_0(R);R]}{\partial{\Delta_0}^2} D_2(R)^2 + \dots \right\} \;.
\end{equation}
We find the elliptical symmetric part
$\Delta_0(R)=D_0(R)/R\sqrt{4\pi}$ of the gap by neglecting the
$D_2$ contribution and minimizing the thermodynamic potential
with respect to $\Delta_0(R)$. This gives a spherical symmetric
equation similar to \eqref{eq:TTI:gapEOM}, but now with a
slightly smaller gradient coefficient, given by $\gamma_0$ in
\eqref{eq:TTI:gradienCoefs}. When we have obtained a solution
for $D_0$ we can minimize the thermodynamic potential with
respect to $D_2$, which gives the following linear-response
equation
\begin{align}
    {\cal L}  D_2(R) & =
    S \left(D_0(R);R\right)\;,
\end{align}
with the linear operator
\begin{align}
    {\cal L} & = \frac{1}{2}\frac{\partial^2\omega_{\rm BCS}[\Delta_0(R);R]}{\partial{\Delta_0}^2} - \frac{\gamma_2}{2}\left(\frac{d^2}{dR^2} -
    \frac{6}{R^2}\right)
\end{align}
and the inhomogeneous term that acts as a source for the
quadrupole deformations
\begin{align}
    S \left(D_0(R);R\right) & =
    \frac{\gamma_{02}}{2}\left(\frac{d^2}{dR^2}D_0(R)-\frac{3}{R}\frac{d}{dR}D_0(R)+\frac{3}{2R^2}D_0(R)\right)\;.
\end{align}

\begin{figure}[t]
\centering
\includegraphics[height=4.2cm]{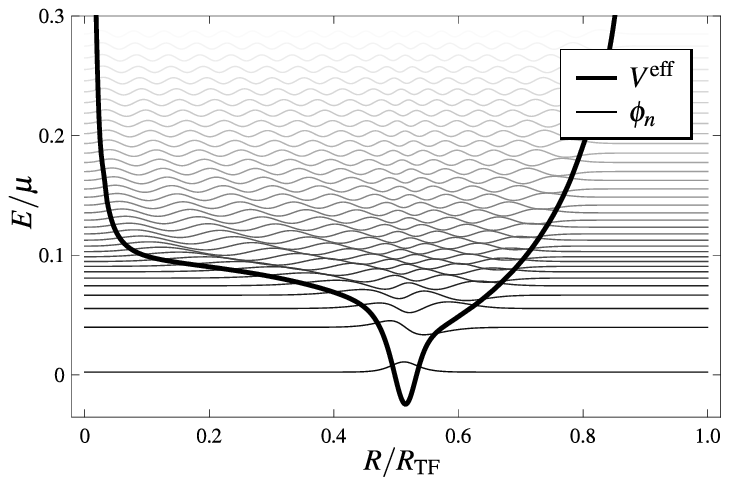}
\hfill
\includegraphics[height=4.2cm]{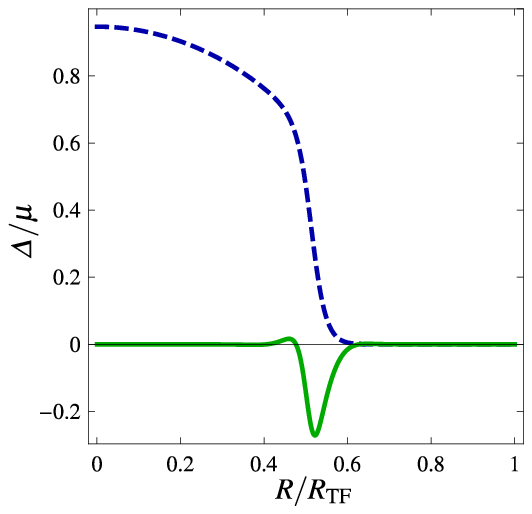}\hspace*{.6cm}
\caption{ The left panel shows the solutions for the
eigenfunctions (thin lines) and eigenvalues (line height) of
\eqref{eq:TTI:EigenEquations}. The thick line is the effective
potential in \eqref{eq:TTI:EigenEquations}, which shows a
pronounced dimple at the location of the interface. The right
panel shows the elliptically symmetric solution $\Delta_0(R)$
(dashed line) and the quadrupole correction $D_2(R)$ (solid line).
Here we have taken $\alpha = 45$ and $P = 0.4$, which are typical
values for the experiments of Partridge {\it et al}.\
\cite{partridge2006dtf}. } \label{fig:TTI:YexpSols}
\end{figure}

In Dirac notation the solution of this equation is formally
given by $|D_2\rangle= {\cal L}^{-1} |S(D_0)\rangle$. Inverting
the operator ${\cal L}$ can be accomplished by first
diagonalizing this operator, which we can do by finding all its
eigenfunctions and eigenvalues. Interestingly, these are
determined by a Schr\"odinger equation
\begin{align}\label{eq:TTI:EigenEquations}
    \left\{-\frac{\hbar^2}{2 m^*}\frac{d^2}{dR^2} + V^{\rm eff}(R)\right\}\phi_n(R) = E_n\phi_n(R)
    \;,
\end{align}
with an effective mass given by $m^*=\hbar^2/\gamma_2$ and an
effective potential $V^{\rm eff}(R)$
\begin{align}\label{eq:TTI:EigenEquations}
  V^{\rm eff}(R) = \frac{1}{2}\frac{\partial^2\omega_{\rm BCS}[\Delta_0(R);R]}{\partial{\Delta_0}^2} +
\frac{\hbar^2}{2m^*}\frac{6}{R^2}\;.
\end{align}
A typical example of this effective potential with its
eigenstates and energies is shown in
Fig.~\ref{fig:TTI:YexpSols}. Given these eigenfunctions the
solution for $D_2$ can in Dirac notation finally be written as
$|D_2\rangle= \sum_n (1/E_n)
|\phi_n\rangle\langle\phi_n|S(D_0)\rangle $, which amounts to
\begin{equation}\label{eq:TTI:sol}
    D_2(R) = \sum_n\frac{\phi_n(R)}{E_n}\int_0^{\infty}\!\dd R'\ \phi_n(R') S\left(D_0(R');R'\right)\;.
\end{equation}
In Fig.~\ref{fig:TTI:YexpSols} also the corresponding solution
for $D_2$ is shown. This solution is centered around the
interface and is also roughly of the same width as the
interface. This is as expected, since the terms in the
thermodynamic potential that do not obey the elliptical
symmetry and are the source for the quadrupole deformations,
are most significant near the interface. Formally, this comes
about because the sum in the right-hand side of
\eqref{eq:TTI:sol} is, due to the energy denominator, dominated
by the eigenstates with low energies that are localized in the
dimple of the effective potential $V^{\rm eff}(R)$.

The outcome of our linear-response analysis gives only rise to
small deformations from the elliptical symmetry. In fact, this
{\it a posteriori} makes this approach self-consistent and
confirms the assumption that the gap can be well described with
a solution that has the same symmetry as the trap. While this
symmetric solution gives roughly speaking the average shape of
the interface, the small quadrupole deformations correct for
this and widen the interface in the radial direction and shrink
it in axial direction. This effect becomes bigger for larger
aspect ratios, but never gives rise to such large deformations
as is seen in the experiments of Partridge {\it et al}. For an
aspect ratio of one, the deformation disappears, because the
source term $S\left(\Delta_0(R),R\right)$ is proportional to
$\gamma_{02}$, which becomes zero at $\alpha=1$. In principle,
a deformation could then occur spontaneously, if one or more
eigenvalues $E_n$ become negative. However, for typical
experimental parameters, this never happens.

In this subsection we discussed the linear response of the
superfluid-normal interface shape. This is a nice application
for our Landau-Ginzburg-like thermodynamic potential functional
that can be used to study in detail the effect of the aspect
ratio of the trap on the experiment of Shin {\it et al}.
However, we cannot use it to describe the large deformations
observed by Partridge {\it et al}. A possible way to handle
this situation requires beyond linear-response methods that are
covered in the next section.

\subsubsection{Metastable States}

In the previous section we assumed that the deviations from the
elliptically symmetric solution for the gap are small and
therefore validates the use of linear response. But since we
have the full thermodynamic potential at our disposal we can
also consider large deviations by using a variational approach.
In the experiment of Partridge {\it et al}.\ the observed
deformation of the superfluid core is indeed large. This
deformation can be modelled by giving the superfluid core a
different aspect ratio than that of the trap
\cite{haque2007tfc} and by letting the polarized normal shell
follow the shape of the trap. It is still unclear whether this
represents the true energy minimum of the system or corresponds
to a metastable state \cite{baksmaty2010cms}. We can use our
thermodynamic potential to investigate this, and we will see
that there appears to be no metastable state in the
Landau-Ginzburg-like theory presented in this chapter.

The superfluid core is described by a nonzero gap function,
which is determined by minimizing the thermodynamic potential.
The case of a metastable state then corresponds to a local, but
not a global minimum of the thermodynamic potential. We want to
find such minima by using a variational approach. This implies
that we somehow have to parameterize a likely functional form
of the gap, and then vary the thermodynamic potential with
respect to these parameters. To find a appropriate trial
function that describes the gap well, let us start with the
following function that very accurately describes the gap in
the elliptically symmetric case
\begin{align}\label{eq:TTI:varGap}
    \Delta_0(R) = \Delta_0  \left(1-\frac{R^2}{\rho R_{\rm TF}^2}\right)
      \frac{\tanh\left(\frac{R_0-R}{\Delta R}\right)+1}{2}\;.
\end{align}
Here $\rho$, $R_0$ and $\Delta R$ are variational parameters.
These parameters can be understood as follows. In the
homogeneous theory the gap is proportional to the chemical
potential $\Delta_0=\zeta\mu \simeq 0.97 \mu$, as discussed
before, and in the trap the chemical potential is given by $\mu
- V({\bf x})
\equiv \mu(1-R^2/R_{\rm TF}^2)$. This explains the first factor in
the right-hand side of \eqref{eq:TTI:varGap}, where the
parameter $\rho$ is needed to incorporate beyond-LDA effects.
The function $[\tanh((R_0-R)/\Delta R)+1]/2$ with center $R_0$
and width $\Delta R$ describes the interface profile, since
this is approximately equal to the usual soliton solution for
an interface in Landau-Ginzburg theory. For specific
temperatures and polarizations a minimum of the thermodynamic
potential with respect to these variational parameters can
easily be found numerically.

\begin{figure}[t]
\centering
\includegraphics{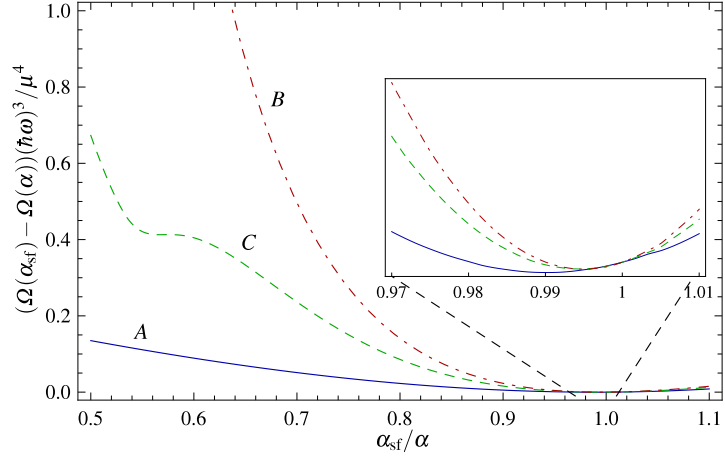}
\caption{%
The total thermodynamic potential of the system as a function of
the deformation $\alpha_{\rm sf}/\alpha$ of the superfluid core.
The different lines correspond to different choices for the
deformation of the superfluid core as discussed in the text. The
solid line shows a simple change in aspect ratio for the
superfluid core only, as in \eqref{eq:TTI:SuperfluidARchange}.
This corresponds to a change in the axial direction only. For the
dashed and dash-dotted line the gap profile also changes in the
radial direction. For an appropriate scaling of the thermodynamic
potential we have introduced the radial trap frequency $\omega$.}
\label{fig:TTI:gapDeformation}
\end{figure}

Let us now also include the aspect ratio in this variational
approach. We want to see how the thermodynamic potential
changes when the superfluid core has a smaller aspect ratio
than the normal shell. Since we consider this in a variational
manner, we need a proper function with a parameter to describe
this. Let us first simply vary the aspect ratio of the gap
profile. This can be achieved by performing in
\eqref{eq:TTI:varGap} the substitution $R \rightarrow R_{\rm
sf}$, with $R_{\rm sf}$ the scaled coordinate of
\eqref{eq:TTI:ellipticalSymmetry} with aspect ratio
$\alpha_{\text{sf}}$. This then results in
\begin{align}\label{eq:TTI:SuperfluidARchange}
    \Omega(\alpha_{\rm sf}) = \int\!\dd\bm x \left(
    \frac{1}{2}\gamma({\bf x})|\nabla\Delta_0(R_{\rm sf})|^2
    + \omega_{\rm BCS}[\Delta_0(R_{\rm sf});{\bf x}]
    \right)\;.
\end{align}
In Fig.\ref{fig:TTI:gapDeformation} the solid line (curve
\emph{A}) shows the total thermodynamic potential as a function
of $\alpha_{\text{sf}}$. For this plot, we choose the trap
aspect ratio to be $\alpha=45$, because this is a typical value
for the experiments of Partridge {\it et al}.\ where
deformation is clearly visible. Also a polarization should be
taken and we choose $P=0.4$ in the elliptically symmetric case
for these figures. The thermodynamic potential, however, does
not show any signs of a dramatic metastable deformation. Yet
the energy minimum is at a slightly smaller aspect ratio for
the superfluid core then the trap. We find that for these
parameters we have $\alpha_{\rm sf} \simeq 0.99 \alpha$. This
very small deformation is consistent with the linear-response
result from the previous subsection.

In the experiment of Partridge {\it et al}., not only the
superfluid-normal interface deforms, but also the partially
polarized shell appears to be absent. To some extent, this can
be reproduced with a gap parameter that is nonzero further to
the outside of the trap in a region where the LDA would predict
it to be zero. Since a nonzero gap forces the system to be
balanced, the majority species will be forced to the outside,
and the gas resembles what is seen in the experiment. In order
to look for a metastable state that does exactly this, we can
parameterize a gap function in different ways. One possibility
(option \emph{B}) is to change the aspect ratio of the gap, not
by shrinking it in the axial direction, but by enlarging it in
the radial direction. This means we replace the radius $R_{\rm
sf}$ in \eqref{eq:TTI:SuperfluidARchange} by
\begin{align}
    (R_{\rm sf})^2 = \left(\frac{\alpha_{\rm sf}}{\alpha}\right)^2(x^2+y^2)+\left(\frac{z}{\alpha}\right)^2\;,
\end{align}
with again $\alpha_{\rm sf}$ the variational parameter that we
can change. An even better option (option \emph{C}) is to
actually shift the location of the interface while changing the
aspect ratio simultaneously. This can be done by using again
$R_{\rm sf}$ as in \eqref{eq:TTI:SuperfluidARchange} and
substituting $R_0
\rightarrow R_0 \alpha/\alpha_{\rm sf}$ in \eqref{eq:TTI:varGap}.

The thermodynamic potential for both options is again plotted
in Fig.~\ref{fig:TTI:gapDeformation}, with the same aspect
ratio $\alpha$ and polarization as for option \emph{A}. The
thermodynamic potential for option \emph{B} has clearly no
features and only one minimum near the elliptically symmetric
solution. The result for option \emph{C}, however, seems to
have a feature that looks like a metastable point near
$\alpha_{\rm sf} = 0.57 \alpha$. A closer look reveals that it
is not a local minimum but a saddle point. This point is a
result of our choice of parametrization, since this is exactly
the point where the center of the interface in
\eqref{eq:TTI:varGap} is equal to the point where the factor
$1-R_{\rm sf}^2/\rho R_{\rm TF}^2$ becomes zero. At this value
of $\alpha_{\rm sf}$ the interface thus disappears.

For the different trial functions of the gap that we considered
here, we can conclude that there is no metastable solution with
a dramatic deformation in this system. There are of course many
more possible trial functions conceivable, but at present it
appears unlikely that any of these contain a clear and deep
enough metastable solution that can explain the dramatic
deformation of Partridge {\it et al}.\ \cite{hulet2006pps}.
Because of the large deformations that we are looking for,
higher-order gradient effects in the gap, or even density
gradient effects, may be very important. We can therefore not
conclude that we should reject metastability as the solution to
this outstanding problem, but it remains a challenge to find
such metastable solutions in a theory that is simultaneously
also able to accurately describe the experiments of Shin {\it
et al}.

\section{Conclusions}

In this chapter we discussed a Landau-Ginzburg-like approach to
the unitarity Fermi gas problem, that we believe is both simple
and elegant. This approach is based on the existence of an
exact thermodynamic potential functional. By taking the most
important interaction contributions into account, we showed
that in this way all known thermodynamic properties of the
homogeneous imbalanced Fermi mixture can be accounted for. When
also the gradient energy of the gap is incorporated, the theory
can be extended to describe inhomogeneity effects of a Fermi
mixture trapped in an external potential in a manner that goes
beyond the usual local-density approximation.

We showed in the first part of this chapter that the
interactions can be incorporated in two frequency and momentum
independent self-energies. We showed that these self-energies
naturally depend on the superfluid gap. The topology of the
phase diagram of the unitary Fermi mixture is correctly
captured by the mean-field BCS-theory. The self-energy
corrections do not change this topology, but change the
critical lines in the phase diagram quantitatively. The results
from experiments and various Monte-Carlo calculations uniquely
determine the two parameters in the self-energy. This results
in a parameter free thermodynamic potential that contains all
known features and has the correct energies and equation of
state for the homogeneous Fermi mixture.

The homogeneous result can be used in a local-density
approximation. To go beyond this approximation, the energy cost
of gradients in the gap needs to be taken into account. With
this additional contribution to the thermodynamic potential we
can describe the superfluid-normal interface in more detail.
The experimental data from Shin {\it et al}.\
\cite{shin2008pdt}, which shows a rather smooth interface, is
very well explained in this way. The smooth interface leads
also to a local stabilization of a gapless superfluid, the
Sarma phase. This interesting prediction of the theory,
however, still needs to be corroborated by further experiments.
The surface tension of the interface can be calculated and
turns out to be rather small. This is consistent with the
observation of Shin {\it et al}., who see experimentally no
deformation of the superfluid core, but it is in sharp contrast
with the observations of Partridge {\it et al}.\
\cite{partridge2006dtf}, who see a dramatic deformation. This
deformation actually suggests a much larger surface tension,
but another explanation may be that in their case the system is
in a metastable minimum of the thermodynamic potential. In a
variational approach we showed, however, that the
Landau-Ginzburg-like model derived in this chapter, most likely
does not contains such a local minimum. Because the deformation
is large, higher-order gradient effects in the gap, or even
density gradient effects, may be very important. These effects
are more complicated to include in the thermodynamic potential,
and are beyond the scope of this chapter.

\end{document}